\documentclass[aps,pre,twocolumn,groupedaddress,showpacs]{revtex4-1}

\usepackage{amsmath}
\usepackage{graphicx}
\usepackage[usenames]{color}

\newcommand{\f}{\bar{f}}

\begin{document}

\title[An Itinerant Oscillator model with cage inertia]{An
Itinerant Oscillator model with cage inertia  for mesorheological granular experiments}

\author{Antonio Lasanta }
\affiliation{CNR-ISC and 
Dipartimento di Fisica, Universit\`a La Sapienza, p.le A. Moro 2, 00185 Rome, Italy}

\author{Andrea Puglisi}
\affiliation{CNR-ISC and 
Dipartimento di Fisica, Universit\`a La Sapienza, p.le A. Moro 2, 00185 Rome, Italy}

\vspace{10pt}

\begin{abstract}
Recent experiments with a rotating probe immersed in weakly fluidized
granular materials show a complex behavior on a wide range of
timescales. Viscous-like relaxation at high frequency is accompanied
by an almost harmonic dynamical trapping at intermediate times, with
possibly anomalous long time behavior in the form of super-diffusion.
Inspired by the Itinerant Oscillator model for diffusion in molecular
liquids, and other models with coupled thermostats acting at different timescales,  here we discuss a new model able to account for fast viscous
relaxation, dynamical trapping and super-diffusion at long times. The
main difference with respect to liquids,  is a non-negligible
cage inertia for the surrounding (granular) fluid, which
allows it to sustain a slow but persistent motion for long times.  The
computed velocity power density spectra and mean-squared displacement
qualitatively reproduce the experimental findings. We also discuss the
linear response to external perturbations and the tail of the
distribution of persistency time, which is associated with
superdiffusion, and whose cut-off time is determined by cage
inertia.
\end{abstract}

%
%
%
%
%

\maketitle

\section{Introduction.}

Granular materials share analogies with condensed ``molecular''
matter, but often escape its well-established theoretical
approaches~\cite{JNB96,BPS15}. Equilibrium statistical physics may
suggest only very approximate ideas about the qualitative behavior of
granular media in strongly fluidized steady states and dramatically
fails in the extreme case of static or quasi-static regimes. Continuum
descriptions for dense flows lack first-principle constitutive
relations~\cite{andreotti,luding}, while more refined kinetic theories
(e.g. mode-coupling) must be carefully adapted to take into account
some fundamental peculiarities, such as the breaking of time-reversal
invariance and the relevance of the inertia of the
medium~\cite{zippelius}. The building up of granular hydrodynamics
from ``microscopic'' models (ie. where all grains are described) is a
promising program, but for the moment its success is limited to dilute
regimes and rests upon the (uncertain) separation between fast and
slow scales~\cite{BP04,puglisi,lasanta12}.

Our understanding of the liquid state of granular matter, being in the
middle between two opposite worlds (the very fast ``granular gases''
and the very slow ``granular glasses''), is even more incomplete and
may benefit from simplified effective models.  An important insight is
provided by experiments, where the ``liquid'' state is realized
through the application of some mild shaking leading to a slowly
mixing flow with strong correlations and long but finite relaxation
times~\cite{danna,hecke2,camille}. When the longest relaxation time
overcomes the experimental times, one may say to have reached a {\em
  transition point}, entering into a sort of - empirically defined -
solid or glassy state~\cite{dyre,barrat,bouchaud}. We do not intend to
directly address such a transition: however some of our results, in
the following, concern also this delicate point.

The present paper aims at discussing a simplified {\em linear} model
which is able to reproduce some noticeable phenomena observed in a 
granular liquid state~\cite{camille}. In particular our ambition is to propose a
minimal model which exhibits a transient cage effect and
super-diffusion at later times. Caging is a common hallmark of
diffusion in dense liquids~\cite{cavagna} and it is usually found in
granular systems at large packing
fractions~\cite{marty,reis2007,zippelius}. Superdiffusion is much less
common in liquids, it seems rather a peculiar effect of granular
systems~\cite{barrat,bouchaud}, however it is rarely seen and hardly
explained: below the jamming transition it has been observed
in~\cite{bouchaud}, above such a transition it was seen
in~\cite{barrat} where it was imputed to ``zero''-modes of the host
fluid, or in~\cite{behringer} where the mechanism of Taylor dispersion
was involved, or in ~\cite{roux} explained by a turbulence-like
cascade effect. A universal scenario for anomalous diffusion is
lacking~\cite{klages}, but certainly it is the signal of an enduring
memory. A family of phenomenological models for anomalous diffusion
includes fractional Fokker-Planck equations~\cite{klafter}, where an
immediate physical interpretation is not always at hand. The
observations in~\cite{camille} were better explained through a
phenomenological continuous time random walk model for the {\em
  velocity}~\cite{castiglione}, with a power-law-decaying distribution of
persistency times which was confirmed by experimental
measurements. Such a model however could not explain the cage effect
(which is a sub-diffusive behavior at earlier times) and was,
therefore, adopted to match only partially the experimental results,
in particular the slow time-scales. The model presented here, on the
contrary, aims at offering a unifying picture for the two phenomena, and
highlights the essential role of the ``cage inertia'', which is
the origin of long-time memory.

The organisation of the paper is as follows. In Section II we
summarize the results of a recent experiment, carried on by one of
the authors, where time correlations and mean squared displacement of
a probe were measured, interpreted under the light of a first
simplified model. In Section III we propose the new Itinerant
Oscillator model with cage inertia, with a discussion of its
motivation. In Section IV we report the main formula for static
quantities in the steady state. In Section V, we discuss two-time
quantities in the steady (time-translantion invariant) state,
including the velocity spectrum, the mean squared displacement, linear
response and, computed only numerically, the distribution of
persistency times. Conclusions and perspectives are drawn in Section
VI.

\section{A recent experiment on granular mesorheology}

\begin{figure}
\includegraphics[width=4cm,clip=true]{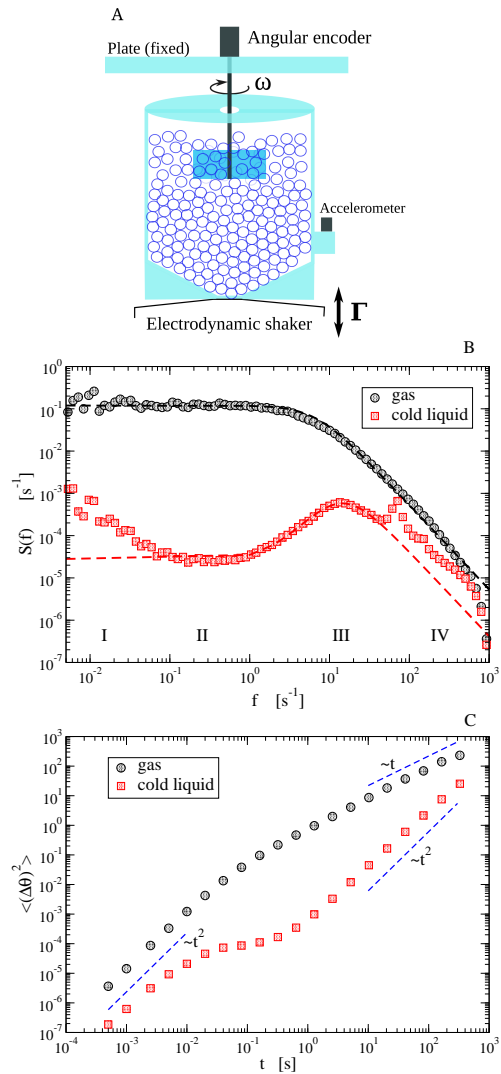}
\includegraphics[width=6.5cm,clip=true]{spe.eps}
\includegraphics[width=6.5cm,clip=true]{dif.eps}
\caption{A: sketch of the experiment reported in~\cite{camille}. B: experimental
  data of the vpds for the gas case and the ``cold liquid'' case,
  together with predictions (dashed lines) from the incomplete
  model, Eq.~\eqref{dhc}. C: experimental data of the msd for both cases,
  together with dashed lines useful as guides for the
  eye.\label{fig:sketch} }
\end{figure}

The granular liquid state is characterized by the emergence of many
time-scales, associated with the complex and collective relaxation
behavior of grains. A window into those time-scales may be open by
studying the dynamics of a diffusing impurity, both in experiments and
in simulations \cite{sar10,camille}. A recent experimental
study~\cite{camille} has offered a new picture in a wide range of
time-scales, from $10^{-3}$ s up to $10^3$ s and more, revealing a non
trivial scenario. In the experimental setup, sketched in
Fig.~\ref{fig:sketch}A the ``impurity'' was constituted by an
immeresed blade (with momentum of inertia $I$) who could rotate around a fixed vertical axis under
the kicks from the grain of a vibrofluidized granular medium. The
dynamics of the angular velocity $\omega(t)$ of the blade and its
absolute angular position $\theta(t)=\int_0^t ds \omega(s)$, was
studied in different regimes of density and intensity of vibration. In
Fig.~\ref{fig:sketch}B, the velocity power density spectrum (vpds)
$S(f)=\frac{1}{2\pi t_{TOT}}\lvert\int_0^{t_{TOT}} \omega(t) e^{i
  (2\pi f) t}dt \rvert^2 $ is presented and its salient features are
highlighted in two opposite limits, which are the gas and the cold
liquid. We remind that the vpds is the Fourier transform of the
velocity autocorrelation function (vacf) and that its $f \to 0^+$
limit is the self-diffusion coefficient, i.e. $D_\infty=\pi \lim_{f\to
  0^+} S(f)$. We also recall that relations exist, under certain
approximations, between the vpds and the intermediate scattering
function which - in liquids - is typically accessed through neutron
scattering experiments \cite{rahman}.

In the gas limit, when the packing fraction is low and the average
energy per grain is high, the probe velocity autocorrelation (vacf) is
close to a simple exponential decay $\sim e^{-t/\tau_{gas}}$, ruled by
a single relaxation time $\tau_{gas}$: in this limit the vpds takes
the form of a Lorentzian $S(f) \propto \frac{1}{\pi \gamma}/[1+(2\pi
  If/\gamma)^2]$ with $\gamma=I/\tau_{gas}$.

In the cold liquid limit, when the packing fraction is high (larger
than $30-35\%$) and the average energy per grain is small (but still in
an ``ergodic'' phase), the observed vpds strongly deviates from the
Lorentzian. Ignoring a mechanical resonance due to the mounting plate
at $\sim 10^2 Hz$, it displays four different regions: at high
frequency (region IV) it decays with a negative power law equal or
smaller than $2$; in region III it shows a smooth parabolic maximum
(centered near $\sim 10 Hz$, reminiscent of a harmonic confinement
(``cage''); in region II it stabilizes on a short plateau, which
suggests a loss of memory (as in the plateau of the Lorentzian which
marks the onset of normal diffusion); finally region I, perhaps the
most surprising one, shows an increasing $S(f)$ for $f \to 0^+$,
signaling a problem with the finiteness of the self-diffusion
coefficient $D_\infty$. A few longer experiments (12 hours) were
conducted, showing a slow crossover toward a new higher plateau at
very low frequencies. The study of the mean squared displacement
(msd), see Fig.~\ref{fig:sketch}C confirmed that the four regions of
the cold liquid case correspond, respectively, to short-time ballistic
(free) motion (IV), dynamical arrest due to caging (III), later
relaxation of the cage (II) and ``final'' superdiffusive behavior (I).

In~\cite{camille} a first model was proposed to account for regions
II-III-IV of the vpds in the cold liquid limit. The model, which was
inspired by a model for diffusion in cold liquids called Itinerant
Oscillator model~\cite{io1,io2,vollm}, also recently used to describe
microrheology in living matter~\cite{FKHVW14}, describes the evolution
in time of the angular velocity of the probe $\omega(t)$ and its
angular position $\theta(t)=\int_0^t ds \omega(s)$, according to the
following stochastic equations of motion:
\begin{subequations} \label{dhc}
\begin{align}
I\dot{\omega}(t)&=-\gamma\omega(t)-k[\theta(t)-\theta_0(t)]+\sqrt{2 \gamma T}\eta(t),\\
\theta(t)&=\int_0^t \omega(s)\\
\theta_0(t)&=\sqrt{2D_0}\int_0^t\eta_0(s)\\
\end{align}
\end{subequations}
where $\eta(t)$ and $\eta_0(t)$ are independent white normal Gaussian
noises (unitary variance). The model represents the diffusion of a
particle in a harmonic potential with ``stiffness'' $k$ and unfixed
minimum located at $\theta_0(t)$, under the effect of a thermal bath
at temperature $T$ and relaxation time $I/\gamma$. The harmonic
potential, representing the cage created by the confining effect of
the dense granular host fluid, is not fixed but moves, as
$\theta_0(t)$ behaves as Brownian motion with diffusivity
$D_0$. Motivation for this model is twofold: 1) it reproduces the main
features of the vpds, i.e. short time fast relaxation (region IV), an
elastic resonance at intermediate times (region III) and a plateau
revealing loss of memory at larger times (region II); 2) in the dilute
limit (when $k \to 0$) it can be rigorously derived~\cite{bromo},
3) at intermediate densities a series of studies showed that memory
effects (coming from correlated collisions) are well described by a
similar coupling with an additional degree of freedom characterized by
slower relaxation time-scales~\cite{sar10}.  The vpds of the above
model can be calculated and reads
\begin{equation} \label{sp_dhc}
S(f) = \frac{1}{\pi}\frac{D_0 k^2+\gamma T (2\pi f)^2}{\gamma^2 (2\pi f)^2+[k-I(2\pi f)^2]^2}~~.
\end{equation}
Two limiting cases are recovered: when $k=0$, the Ornstein-Uhlenbeck
process is obtained, with $S(f)$ taking the Lorentzian form mentioned
before. When $k>0$ and $D_0=0$, one has the Klein-Kramers process in a
fixed harmonic potential, and $S(f) \to 0$ for $f \to 0$, expressing
the absence of diffusion at large times: the cage does not move and
fully confines the particle.  Formula~\eqref{sp_dhc} fairly fits the
experimental spectra (see dashed lines in Fig.~\ref{fig:sketch}B) in
regions II-IV, with $k/I \sim (2\pi \cdot 10)^2 \sim 4 \cdot 10^3
Hz^2$ see ~\cite{camille}.  Reasonably, the ``cage stiffness''
decreases at increasing shaking intensity. It also decreases as the
density is reduced, and abruptly goes to zero at packing fractions of
the order of $\sim 15\%$. The ``cage diffusivity'' $D_0$ rapidly
increases with increasing $\Gamma$ and with decreasing packing
fraction (or number of particles).

The main problem of model~\eqref{dhc} is that in region I it always
predicts a diffusive behavior, no super-diffusion is
allowed. In~\cite{camille} superdiffusion was put in strict relation
with a broad distribution of ``inversion'', also called
``persistency'', times measured with the following recipe. The
frequencies of regions III and IV were filtered out, by considering  a
smoothed angular velocity $\omega_s(t)=\frac{1}{\tau}\int_0^\tau \omega(s)ds$ ($\tau> 1
$s), which displayed much smaller fluctuations than those of
$\omega(t)$, but with positive correlation for very long times (larger
than $10$ s). This behavior, incompatible with model~\eqref{dhc}
(whose relaxation times are much smaller), is likely to be due to a
quasi-steady motion of a large part of the granular medium
surrounding and therefore dragging the probe. The large inertia is
rensponsible for the observed long relaxation times. At high
frequencies (part of region II, region III and IV) the contribution of $\omega_s(t)$ 
 is negligible, explaining the good agreement with model~\eqref{dhc} in that
part of the spectrum.

In the following Section we propose an extension of this model, taking
account the inertia of the surrounding medium, in order to reproduce
the superdiffusive behavior.

\section{The Itinerant Oscillator Model with cage inertia}

Motivated by the experimental measurement of long relaxation times, we introduce a new model to replace (or extend) that of Eq.~\eqref{dhc}:
\begin{subequations} \label{model}
\begin{align}  
I \dot{\omega}(t)&=-\gamma \omega(t) -k[ \theta(t) - \theta_{0}(t)]+ \sqrt{2\gamma T}\eta(t)\\ 
I_{0}\dot{\omega}_{0}(t)&=-\gamma_{0} \omega_{0}(t) +k[ \theta(t) - \theta_{0}(t)]+ \sqrt{2\gamma_0 T_0}\eta_{0}(t) \label{model2} \\
\theta(t)&=\int^{t}_{0}\omega(t')dt' \label{model3}\\ 
\theta_{0}(t)&=\int^{t}_{0}\omega_{0}(t')dt'. \label{model4}
\end{align}
\end{subequations}
In the above equations $\eta(t)$ and $\eta_{0}(t)$ are independent
white gaussian noises with zero average and unitary variance, namely
$\langle \eta(t) \eta(t') \rangle= \delta (t-t')$ and $\langle
\eta_{0}(t) \eta(t') \rangle= \delta (t-t')$.

In equations~\eqref{model} the angular velocity of the probe feels 
two different forces: of course they are both related to collisions,
but one part is without memory and is described by the
Ornstein-Uhlenbeck contribution $-\gamma\omega+\eta(t)$, while the
second part takes the form $-k[ \theta(t) - \theta_{0}(t)]$ and
therefore depends upon the past history of $\omega(t)$ and
$\omega_0(t)$. The choice of a harmonic interaction aims at simplifying
the computations and can be justified by the small velocities of the blade
with respect to that of the surrounding particles. The delaying force
is modelled as a drag toward a reference point which slowly evolves in
time, according to Eqs.~\eqref{model2}-\eqref{model4}. The variable
$\theta_0(t)$ should be viewed as a collective degree of freedom
representing the preferential point of the blade with respect to some
granular {\em cage}. The {\em cage} slowly changes its configuration
and favours the blade's drift at later times. 

In the previous, simpler, model, Eq.~\eqref{dhc}, the dynamics of
$\theta_0(t)$ was {\em overdamped}, as in a motion without
inertia. The new model takes into account the crucial effect of cage
inertia, through the introduction of a fourth degree of freedom
$\omega_0(t)=\dot{\theta}_0$ which evolves with Eq.~\eqref{model2}, as
well as the effect of the blade upon the granular material through the
reciprocal elastic drag. This last ingredient is likely to be
negligible, in view of the large value of the inertia $I_0$, but its
inclusion is convenient for symmetry and physical
consistency. We must emphasize that this model can be
  understood in the context of other models, often referred to as
  ``two temperature models'' (see \cite{Cugliandolo2000,soarez} and
  references therein), where a particle moves under the influence of
  different thermostats acting at different timescales. Even if it is
  quite natural, in dissipative systems such as a granular fluid, to
  associate different temperatures to different timescales, in the
  following we show that the main subject of our study, which is
  super-diffusion, can be obtained even for equal temperatures.

Introduction of cage inertia is the main novelty with respect to the
original Itinerant Oscillator model, including the version discussed
in~\cite{camille}, and certainly deserves some motivation. At low
temperatures, activated processes (that is the possibility to ``jump
out'' of the cage thanks to some thermal fluctuation) are negligible
and the probe never really escapes from a cage: on the contrary, it is
the cage that slowly evolves and dictates the motion of the probe at
large times.  A displacement of the probe $\Delta \theta(t)$ between
two instants separated by a large time $t$, therefore, is closely
related to a displacement $\Delta \theta_0(t)$ of the cage itself,
i.e. of a large part of the surrounding granular medium. Then it is
reasonable, when looking for an ingredient reproducing almost
ballistic superdiffusion, to immagine that the cage is doing long
ballistic drifts (at very small velocity), sustained by its large
inertia. Discreteness and finiteness of the granular material, which
in the experiments is made of a few thousands grains, makes random but
persistent (also called ``secular''~\cite{PADCC14}) drifts
possible. For those reasons it seems reasonable to us the introduction
of a large ``cage inertia'' $I_0$. The event of a large portion of
solvent to drift in a particular direction for times larger than the
cage relaxation is highly unlikely in a dense molecular liquid: this
explains why one usually do not observe superdiffusion in ordinary
liquids.  It is out of the scope of the present paper to discuss the
exact mechanism of formation of the large cage inertia $I_0$ in a
vibrated granular medium, as well as the reason why it increases when
the vibration amplitude is reduced \cite{CFVV99}. The answers to such questions are
postponed to future investigations: our main aim, here, is to convince
the reader that the concept of ``cage inertia'' is useful for the
description of low temperature granular liquids.

The model is linear and it is useful to recast it into a more compact form, by defining
the vector $w(t)=\{z(t),\omega(t),\omega_0(t)\}$, with
$z(t)=\theta(t)-\theta_0(t)$. The model then takes the form
\begin{equation} \label{mod2}
\dot{w}=- A w+ B \tilde{\eta}(t),
\end{equation}
with
\begin{eqnarray}\label{matrdefini}
&&A=\left( {\begin{array}{ccc}
    0 & -1 & 1\\
   \frac{k}{I} & \frac{\gamma}{I}  & 0\\
   \frac{-k}{I_{0}}& 0 & \frac{\gamma_{0}}{I_{0}} \\
  \end{array} } \right), \\
&&B=\left( {\begin{array}{ccc}
    0 & 0 & 0\\
   0 & \frac{\sqrt{2\gamma T}}{I}  & 0\\
   0& 0 & \frac{\sqrt{2\gamma_{0} T_0}}{I_{0}} \\
  \end{array} } \right), \\
&&\tilde{\eta}=\left( {\begin{array}{c}
    0\\
   \eta(t)\\
   \eta_{0}(t)\\
  \end{array} } \right).
\end{eqnarray}

The analysis
of the model will be discussed focusing on a few particular cases, inspired by the experimental results of~\cite{camille},
which are listed in Table~\ref{casi}. By fixing $I=1$ and $T=1$, the
units of moment of inertia and time are fixed. A comparison with
experimental observations suggests that our arbitrary
time unit is close to $1$ real second. In Table~\ref{casi}, case A is
similar to a dilute experiment, while cases D or E are similar to a
dense and cold one. Case B is an example of cage without inertia
($I_0=0$), which cannot display superdiffusion. Case C is similar to
that, as $I_0$ is small. Finally Case F has still a large cage inertia, but has the peculiarity to be at
thermodynamic equilibrium, i.e. $T_0=T$. Nevertheless, we will see that
this ingredient is relevant only in the study of Fluctuation-Dissipation
relation and has no consequences for the presence of cages or
superdiffusion, as it does not crucially affect the timescales of relaxation.

\begin{table}
\begin{tabular}{|c|c|c|c|c|c|c|c|}\hline
        & $I$  & $T$  & $\gamma$  & $k$  & $I_0$  & $T_0$  & $\gamma_0$  \\ \hline
case A  & $1$  & $1$  & $100$  & $0$  & $/$  & $/$  & $/$  \\ \hline
case B  & $1$  & $1$  & $100$  & $5000$  & $0$  & $0.1$  & $1000$  \\ \hline
case C  & $1$  & $1$  & $100$  & $5000$  & $10$  & $0.1$  & $1000$  \\ \hline
case D  & $1$  & $1$  & $100$  & $5000$  & $10^3$  & $100$  & $10$  \\ \hline
case E  & $1$  & $1$  & $100$  & $5000$  & $10^4$  & $100$  & $10$  \\ \hline
case F  & $1$  & $1$  & $100$  & $5000$  & $10^4$  & $1$  & $10$  \\ \hline
\end{tabular}
\caption{Values of the parameters for the cases illustrated in the
  paper. A: The Ornstein-Ulenbeck process.  B: A case without cage inertia
  (overdamped cage). C, D, E: cases with small, medium or large
  cage inertia.  F: equal to case E, but at thermodynamic equilibrium $T_0=T$.} \label{casi}
\end{table}

\section{Statics: one time quantities in the stationary state}
\label{statics}

Eqs.~\eqref{mod2} constitute a linear system, which is solved by
Gaussian multivariate distributions. We assume that the values of the
parameter of the model are such that only eigenvalues with positive real part appear
and a steady state can be reached. In such a steady
state, the one time distribution function takes the form
\begin{equation}\label{disprob}
P(z,\omega,\omega_0)=\frac{1}{\sqrt{(2\pi)^3det({\sigma})}} Exp\left[-\frac{1}{2}w^{T}\beta w\right]
\end{equation}
where $\beta=\sigma^{-1}$ and $\sigma$ is the covariance matrix, which we write in the form
\begin{eqnarray}\label{covariance}
&&\sigma=\left( {\begin{array}{ccc}
    \sigma_{zz} & \sigma_{z\omega} & \sigma_{z\omega_{0}}\\
   \sigma_{z\omega} & \sigma_{\omega\omega}  & \sigma_{\omega\omega_{0}}\\
   \sigma_{z\omega_{0}} & \sigma_{\omega\omega_{0}} & \sigma_{\omega_{0}\omega_{0}} \\
  \end{array} } \right).
\end{eqnarray}
The covariance matrix satisfies the equation \cite{gardiner1985handbook}
\begin{equation} \label{sigsta}
A \sigma + \sigma A^{T}=BB,
\end{equation}
which is rewritten as
\begin{subequations} \label{equsig}
\begin{align} 
\sigma_{z\omega}&=\sigma_{z\omega_0}\\ 
I(\sigma_{\omega\omega_0}-\sigma_{\omega\omega})+k\sigma_{zz}+\gamma\sigma_{z\omega}&=0\\
I_0(\sigma_{\omega_0\omega_0}-\sigma_{\omega\omega_0})-k\sigma_{zz}+\gamma_0\sigma_{z\omega_0}&=0\\
 I(k\sigma_{z\omega}+\gamma \sigma_{\omega\omega})&=T\gamma\\ 
I_0(\gamma_0 \sigma_{\omega_0\omega_0}-k\sigma_{z\omega_0})&=T_0\gamma_0\\
\frac{k\sigma_{z\omega_0}+\gamma \sigma_{\omega\omega_0}}{I}+
\frac{\gamma_0 \sigma_{\omega\omega_0}-k\sigma_{z\omega}}{I_0}&=0. 
\end{align}
\end{subequations}
The solution reads
\begin{subequations} \label{covelem} 
\begin{align}
\sigma_{\omega\omega}&=\frac{T}{I}-\gamma_0 (T-T_0) \frac{A_1}{d} \\
\sigma_{\omega_{0}\omega_{0}}&=\frac{T_0}{I_{0}}+\gamma (T-T_0) \frac{A_1}{d} \\ 
\sigma_{zz}&=\frac{T}{k}-(T-T_0)\frac{A_2}{d} \\ 
\sigma_{\omega\omega_{0}}&=(T-T_0) \frac{A_3}{d} \\
\sigma_{z\omega}&=\sigma_{z\omega_{0}}=(T-T_0)\frac{A_4}{d} ,
\end{align}
\end{subequations}
with $A_1=k (\gamma_0 I+\gamma I_0)$, $A_2=\frac{\gamma_0 \left(I^2 k (\gamma +\gamma_0)+\gamma ^2 \gamma_0 I+\gamma ^3 I_0\right)}{k}$, $A_3=\gamma
\gamma_0 k (I-I_0)$, $A_4=\gamma  \gamma_0 (\gamma_0 I+\gamma  I_0)$ and $d=(\gamma +\gamma_0) \left(\gamma_0 I^2 k+\gamma
\gamma_0^2 I+\gamma  I_0^2 k+\gamma ^2 \gamma_0  I_0\right)$.

\begin{figure}
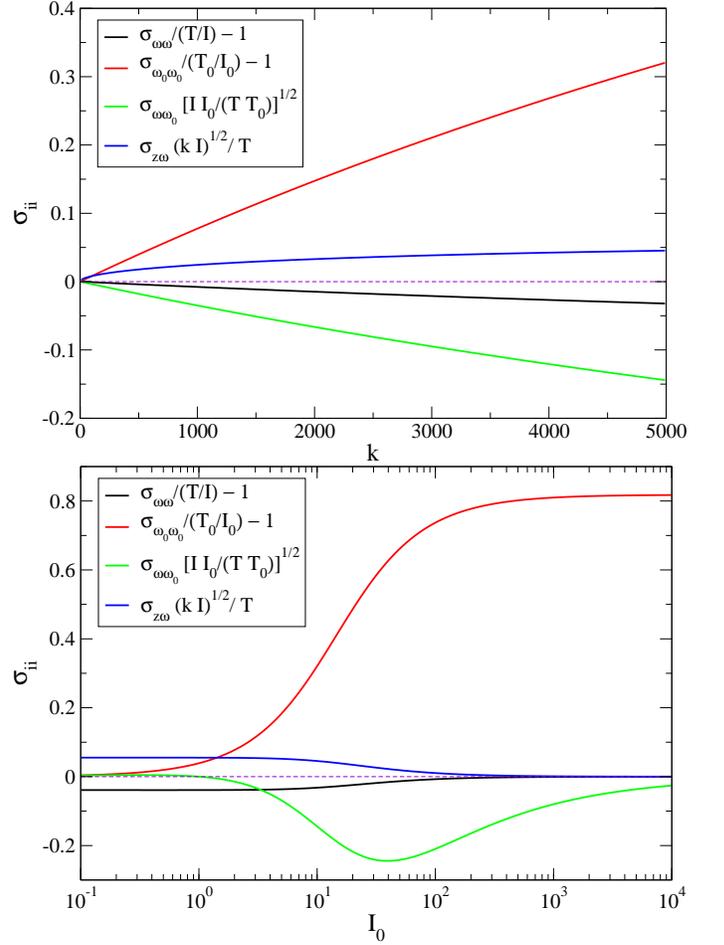

\includegraphics[clip=true,angle=0,width=0.5\textwidth]{sigma_variok.eps}
\includegraphics[clip=true,angle=0,width=0.5\textwidth]{sigma_varioi.eps}
\caption{ We show how $\sigma_{ii}$ (for some $i$, i.e. $\omega$, $\omega_0$ and $z$, properly normalized) depend
  on $k$ or $I_0$, while all the other parameters are fixed according
  to case C of Table~\ref{casi}. On the top panel, we see the
  important effect of the coupling quantity $k$ on both
  self-correlations and cross-correlations. On the bottom one, we
  notice that the effect of cage inertia $I_0$ on the coupling
  (e.g. $\sigma_{\omega\omega_0}$, green curve), is larger for
  intermediate values: at smaller values $\sigma_{\omega_0\omega_0}$
  is close to its ``uncoupled'' value, while it deviates from it at
  large values; the opposite happens to $\sigma_{\omega\omega}$.}
\label{sigma}
\end{figure} 
We can recover two well known physical limits.
In the decoupling limit $k \to 0$, one has
\begin{subequations}\label{eqpl}
\begin{align}
\sigma_{\omega\omega}&= \frac{T}{I}\\ 
\sigma_{\omega_0 \omega_0}&=\frac{T_0}{I_0} \\ 
\sigma_{zz}&= \infty\\ 
 \sigma_{\omega\omega_0}&=0 \\ 
\sigma_{z\omega} &= \sigma_{z\omega_0} =\frac{T-T_{0}}{\gamma+\gamma_{0}}.
\end{align}
\end{subequations}
It is important to note that this limit is singular (since the case
$k=0$, corresponding to two independents Ornstein-Uhlenbeck processes,
is non-stationary for $z$): this explains why
$\sigma_{z\omega}=\sigma_{z\omega_0}\neq 0$.

For $T=T_0$, that is at thermodynamic equilibrium, one has
\begin{subequations}\label{eqpl2}
\begin{align}
\sigma_{\omega\omega}&= \frac{T}{I}\\ 
\sigma_{\omega_0 \omega_0}&=\frac{T}{I_0} \\ 
\sigma_{zz}&= \frac{T}{k} \\ 
\sigma_{\omega\omega_0}&=0. \\ 
\sigma_{z\omega} &= \sigma_{z\omega_0} = 0.
\end{align}
\end{subequations}

In Fig.~\ref{sigma}, we have reproduced the values of the most
relevant covariances as a function of $k$ or $I_0$ in a
case with $T \neq T_0$. The coupling $k$ produces a shift of the two
``temperatures'' $I\sigma_{\omega\omega}$ and
$I_0\sigma_{\omega_0\omega_0}$. It also produces the appearance of cross correlations
$\sigma_{\omega \omega_0}$: it is crucial to note, from
Eq.~\ref{eqpl2}, that such cross correlation is empty in the
equilibrium case, even in the presence of the coupling $k>0$. 

\section{Dynamics}

\subsection{Characteristic times}

\begin{figure}
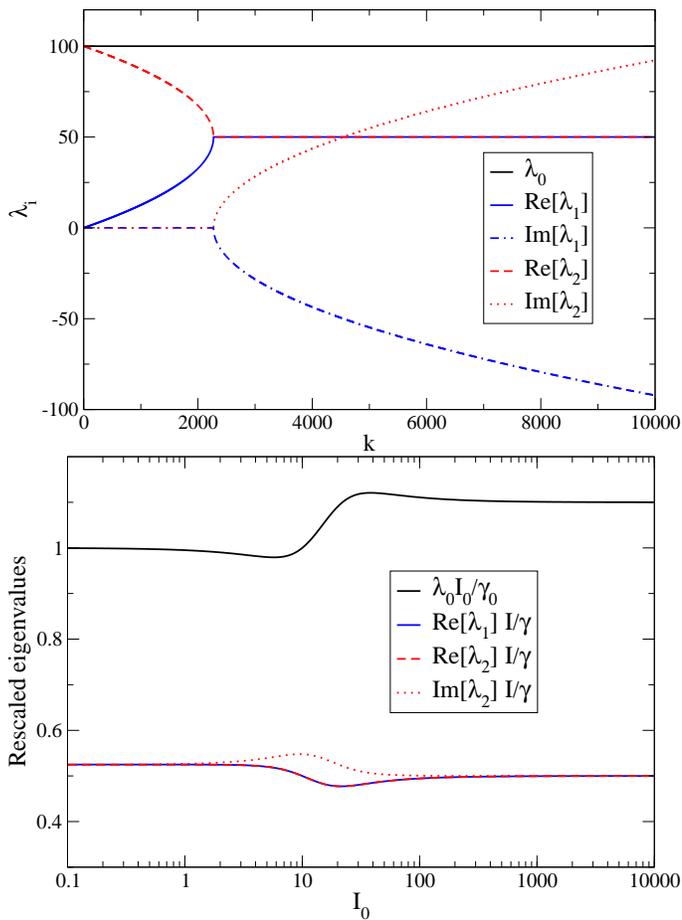

\includegraphics[clip=true,angle=0,width=0.5\textwidth]{ei_variok.eps}
\includegraphics[clip=true,angle=0,width=0.5\textwidth]{ei_varioi0.eps}
\caption{ The three eigenvalues of matrix $A$ as a function of $k$ or
  $I_0$. The other parameters are fixed as in case C of
  Table~\ref{casi}. It is interesting to see that the experimental
  observations in~\cite{camille} occurred in the region beyond the
  bifurcation, i.e. the one with complex eigenvalues.}
\label{eig}
\end{figure} 

The dynamics is characterized by three relaxation times, which are the
inverse of the eigenvalues of matrix $A$. The latter are the zeroes of the
characteristic (cubic) polynomial: a compact
expression is not available in general. The equation for the eigenvalues reads
\begin{equation}
 -\lambda^{3}+B_2\lambda^{2}-B_1\lambda+B_0=0,
\end{equation}
with
\begin{subequations}
\begin{align}
 B_2&=\frac{\gamma_{0} I + \gamma I_{0}}{I I_{0}} \\
 B_1&=\frac{\gamma_{0} \gamma +k(I+I_{0})}{I I_{0}} \\
 B_0&=k\frac{\gamma_{0}+ \gamma}{I I_{0}}.
\end{align}
\end{subequations}
For the solutions to be complex or not, the value of the discriminant must be compared to zero:
\begin{equation}
\Delta =18 B_2 B_1 B_0 -4 B_2^{3} B_0 +B_2^{2}B_1^{2}-4B_1^{3}-27B_0^{2}.
\end{equation}
If $\Delta<0$ there are one real and two complex conjugated
eigenvalues.  If $\Delta\ge 0$ the eigenvalues are all real. Therefore
a bifurcation point appears when $\Delta=0$ and this happen depending
on the values of $\gamma_{0}$, $\gamma$, $k$,$I$,$I_{0}$. We do not
intend to exhaust all the possibilities. In Figure~\ref{eig} we have
reported a plot of the eigenvalues when $k$ or $I_0$ are varied,
keeping constant all the other parameters. The bifurcation when $k$ is
increased is quite evident. It is interesting to notice that a
comparison with experimental data, see~\cite{camille} where $k/I \sim 4 \cdot 10^3$, suggests values of
the parameter in the phase with complex eigenvalues.

It should be noted that, even when all eigenvalues are real, the
correlation functions (discussed in details below) can show
non-monotonic behavior because of the superposition of exponentials
with different characteristic times and with positive/negative
coefficients. The appearance of complex eigenvalues is a source of
persistent oscillatory behavior in the correlation function, which is
of course always damped at large time by the exponential with remaining real
eigenvalue.

\subsection{Velocity power density spectrum}

In this section we study the vpds for the probe's angular velocity $\omega(t)$, which is defined as
\begin{equation}
S(f)=\lim_{t_{TOT}   \rightarrow \infty} \frac{1}{2\pi t_{TOT}}\left|\int^{t_{TOT}}_{0}\omega(t)e^{2\pi fi}dt\right|^{2},
\end{equation}
which is also equivalent to the Fourier transform of the
autocorrelation function in the steady state
$\langle\omega(t)\omega(0)\rangle$ (see next Section for a
discussion). In the following, to avoid confusion with the probe's
angular velocity $\omega(t)$, we will use the symbol $\f=2\pi f$ to
denote the angular frequency associated with $f$, whose use makes more
compact the formula.


By time-Fourier-transforming Eqs.~\eqref{model} and taking the squared modulus, the following formula for the vpds is obtained:
\begin{widetext}
\begin{equation}\label{Sf} 
S(\f)=\frac{1}{\pi}\frac{\gamma T [k^{2}+\f^{2}(\gamma_{0}^{2}-2kI_{0})+\f^{4}I_{0}^{2}]+ \gamma_{0}T_{0}k^{2}}
{k^{2}(\gamma + \gamma_{0})^{2}+\f^{2}\{k[(I+I_{0})^{2}k-2I_{0}\gamma^{2}]+(\gamma^{2}-2Ik) \gamma_{0}^{2}\}+\f^{4}[I_{0}^{2}\gamma^{2}-2II_{0}^{2}k+I^{2}(\gamma_{0}^{2}-2I_{0}k)]+\f^{6}I^{2}I_{0}^{2}}.
\end{equation}
\end{widetext}
The formula is quite rich and may correspond to very different shapes
depending on the choices of the many parameters. We are motivated by a
qualitative comparison with the experimental shapes, see
Fig.~\ref{fig:sketch}B, and for this reason restrict our study to a
few paradigmatic choices of parameters listed in Table~\ref{casi}.

\begin{figure}
\includegraphics[clip=true,angle=0,width=0.5\textwidth]{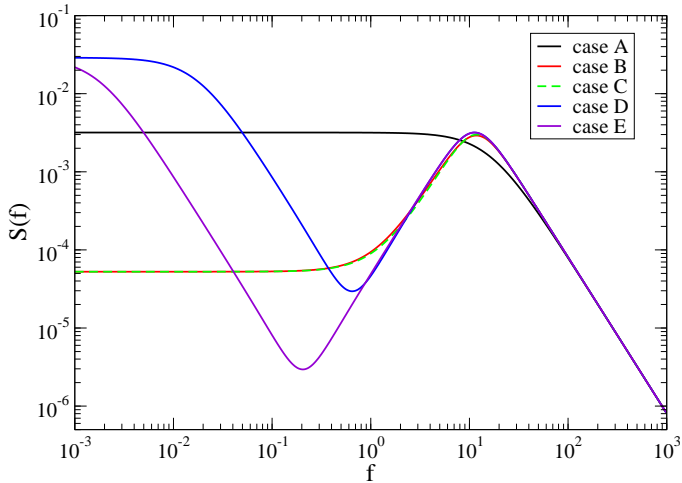}
\caption{ Velocity power density spectrum (vpds) for some cases. The
  parameters of the model are illustrated in Table~\ref{casi}. The
  usual Lorentzian vpds is observed for the uncoupled case A, while for the
  coupling case with zero or small cage inertia (cases B and C) we recover
  the same shape (plateau followed by a bump) as in the ``overdamped model'' discussed in
  \cite{camille}. In cases D and E, that is
  with non negligible cage inertia $I_0$, a region on the left of the resonant bump appears
  with a power law decay resembling experimental observations at small $f$~\cite{camille} .}
\label{psd}
\end{figure}

In Fig.~\ref{psd} the numerical computation of formula~\eqref{Sf} is displayed for such
choices. As expected, case A (no coupling, typical of dilute gases)
reproduces the classical Lorentzian form for the spectrum of the
Ornstein-Uhlenbeck process, as verified by substituting $k=0$ in Eq.~\eqref{Sf} which gives
\begin{equation} \label{lorentz}
S(\f)=\frac{T\gamma}{\pi(\f^{2} I^{2}+\gamma^2)}.
\end{equation}

Case B is similar to the model used
in~\cite{camille}: it shows the resonant bump due to the cage
effect, but is forced to a plateau in the region at low $\f$, because of the
vanishing cage's inertia, i.e. $I_0=0$ (no superdiffusion). This choice corresponds to an ``overdamped'' dynamics for the collective degree of freedom $\theta_0$, and analytically gives
\begin{equation}\label{i0}
S(\f)=\frac{1}{\pi}\frac{k^2(T\gamma+T_0 \gamma_0)+\f^2 T \gamma \gamma_0^2}{k^2(\gamma+\gamma_0)^2+\f^2[I^2 k^2+\gamma_0^2 (\gamma^2-2Ik)]+\f^4 I^2 \gamma_0^2}.
\end{equation}
The expresion is the analogue of the one reported in~\cite{camille},
with the (not crucial) novelty that here $\theta_0(t)$ is also subject to the
reciprocal of the coupling elastic force.

Case C is similar to B, since $I_0$, even if finite, is still small. Case D and E, on
the contrary, exhibit the effect of growing cage inertia and therefore an
increasing behavior as $\f \to 0$. In both cases, necessarily, $S(f)$ becomes flat
at very small $\f$, since all characteristic times - even if
large - are finite. From Fig.~\ref{psd} it is clear that the effect of
$I_0$ is crucial: it rules the larger relaxation time and determines
the duration of ``anomalous'' part of the spectrum.

It is interesting to realize that the ramp at  small values of $\f$ in cases
D and E has a $\sim \f^{-2}$ behavior which resembles the high $\f$
region. This is reasonable as $\omega$, at small frequencies, is
somehow enslaved by the dynamics of $\omega_0$: the latter is, basically, a
 Ornstein-Uhlenbeck process realized at much longer
timescales. This is also the elementary reason why the mean squared
displacement, as discussed below and as observed in some
experiments~\cite{camille}, is close to ballistic.

\begin{figure}
\includegraphics[clip=true,angle=0,width=0.5\textwidth]{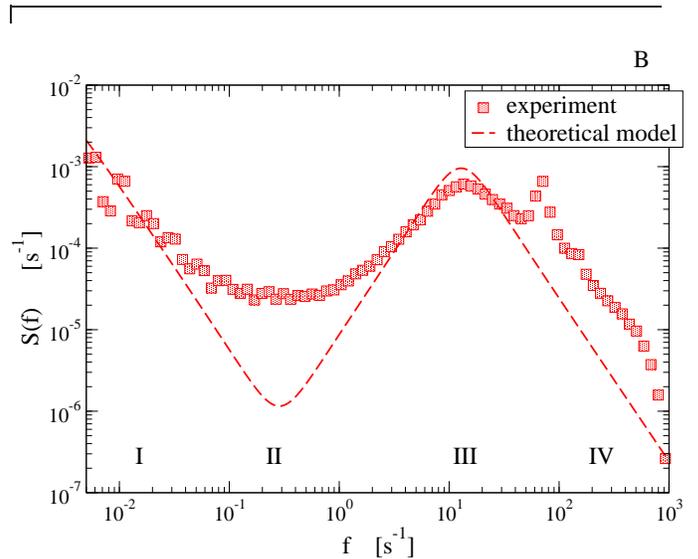}
\caption{ Here, we show a comparison between the theory, Eq.~\eqref{Sf} (dashed lines) and the experimental data (symbols) of the 
velocity power density spectrum (vpds). The parameters of the theory are
$T=0.3$, $I=1$, $\gamma=100$, $k=6500$, $T_0=70$, $I_0=10^4$, $\gamma_0=10$, which are close to case $E$. The experimental data come from Ref.~\cite{camille}, in the cold liquid case.}
\label{psd2}
\end{figure} 
In Fig.~\ref{psd2} we can observe that our model --
even if reproducing, in cases with large cage inertia (D and E), the ramp at small $f$ associated with super-diffusion
-- does not do a perfect job in reproducing the whole shape of the
experimental vpdf.  Indeed, the experimental observations suggest that
the crossover from the cage bump to the ramp at small $f$ is much
smoother, basically flat: our model, on the contrary, predicts
(independently from the parameters) a deep elbow. It
  is no possible to find any region in the parameter space where
  a plateau appears as in the experiment, namely, there is always a
  sharp minimum for the $S(f)$ near the local maximum. Our conjecture
is that, in order to reproduce the smooth crossover, the {\em noise}
$\eta_0(t)$ should be modified, possibly replaced by some coloured
stochastic process. Notwithstanding this discrepancy,
  we will see in the next section that our theory fits very well the
  behavior of the mean square displacement.

\subsection{Autocorrelations and mean squared displacement}

The two-time covariance matrix in steady state, which is time-translational invariance, reads
\begin{equation}
C_{ij}(t)\overset{def}{=}\langle w_i(t)w_j(0) \rangle=\exp(-A t) C_{ij}(0).
\end{equation}
A compact expression of $\exp(-A t)$ is out of question, as already
discussed for the eigenvalues.  We recall that $C_{\omega\omega}(t)
\overset{def}{=}C_{22}(t)=\langle \omega(t)\omega(0) \rangle$ can also
be obtained by the inverse Fourier transform of the vpds. This is true
for each component of $C_{ij}(t)$ which is related - through
transforms - to the cross-spectrum.

\begin{figure}
\includegraphics[clip=true,angle=0,width=0.5\textwidth]{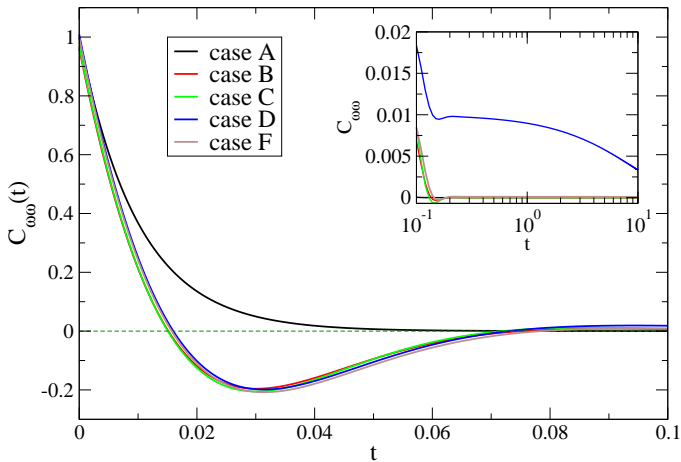}
\caption{ Autocorrelations $C_{\omega\omega}(t)$ for some choices of
  parameters, as illustrated in Table~\ref{casi}. The inset shows the
  same autocorrelations at larger time, in order to highlight the
  persistent behavior of cases with high cage inertia $I_0$ (D and
  F). Interestingly, the cage inertia $I_0$ does not affect in a
  relevant way the behavior of the autocorrelation at time smaller
  than $0.1$.  Case $E$, very close to the $F$ one (different temperatures do not affect the qualitative shape of the autocorrelation), is not shown for the sake
  of clarity}.
\label{autoc}
\end{figure} 

Autocorrelations for some choices of the parameters are numerically
computed (from analytical formula) and shown in Fig.~\ref{autoc}. It
is evident the passage from a simple exponential decay of uncoupled case A ($k=0$) to
the ``back-scattering'' behavior, typical of cages in liquid, in all
other cases with $k \neq 0$. The cage's inertia $I_0$ does not change in a crucial way
the behavior of the first part of the autocorrelation, but determines
a larger and larger persistency of correlations at late times, as seen
in the inset. It is important to realize that the value of the
autocorrelation at large time, even if not vanishing, is very small
with respect to its order of magnitude at small times.

\begin{figure}[b!]
\includegraphics[clip=true,angle=0,width=0.5\textwidth]{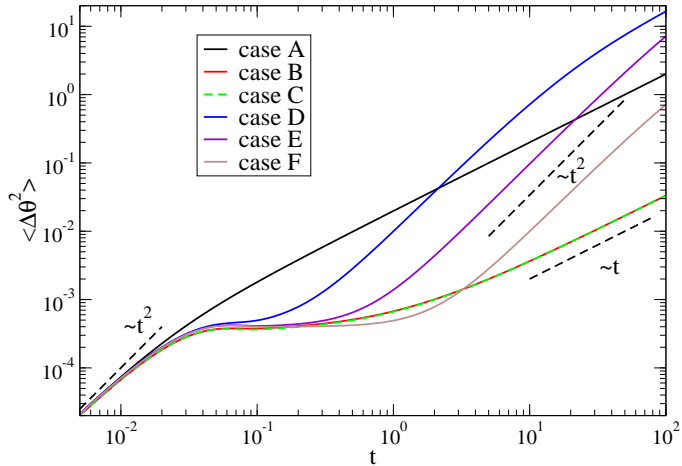}
\caption{ Mean square displacement for some choices of the parameters,
  as illustrated in Table~\ref{casi}. The phenomena observed in the
  vpds are also observed in the MSD. After a common ballistic regimes,
  all cases with coupling (i.e. excluding case A) show an intermediate
  cage effect visible as brief plateau. After the plateau the cases
  with zero or small cage inertia $I_0$ (B and C) show normal
  diffusion, while the cases with large cage inertia (D and E) present
  ballistic super-diffusion. The super-diffusive behavior terminates
  at a time which is larger as $I_0$ increases.}
\label{msd}
\end{figure}

From the knowledge of $C_{\omega\omega}(t)$, it is possible to compute the time-evolution of the mean squared displacement (msd): 
\begin{multline}
\langle [\Delta \theta(t)]^2 \rangle = \int_0^t dt' \int_0^t dt'' \langle \omega(t')\omega(t'') \rangle \\
= 2 \int_0^t dt'(t-t') C_{\omega\omega}(t') .
\end{multline}
In Figure~\ref{msd} we show, for a few choices of the parameters, the
behavior of the msd. The uncoupled case A reproduces the standard
ballistic-diffusive dynamics which is typical of diffusion in dilute
gases. Coupling ($k>0$) induces a dynamical arrest in the form of a
plateau in the msd, which - later - is overcome by the slow cage
dynamics. Such dynamics is purely diffusive in the case without
inertia ($I_0=0$ or small) while it is super-diffusive when $I_0$ is
large. When the observation time is larger than the time dictated by
$I_0$ (basically $I_0/\gamma_0$) the msd comes back to normal
diffusion. In experiments such a very late stage was observed on
timescale of the order of hours, see Supplemental Material
in~\cite{camille}.  We wish to underline that, how anticipated, the
equilibrium case F - which differs from case E just in the fact that
the temperatures are equal ($T=T_0$) - displays superdiffusion as
well. The long memory is induced by the large cage inertia $I_0$ and
is not related to the system being at equilibrium or out of it.

\begin{figure}[b!]
\includegraphics[clip=true,angle=0,width=0.5\textwidth]{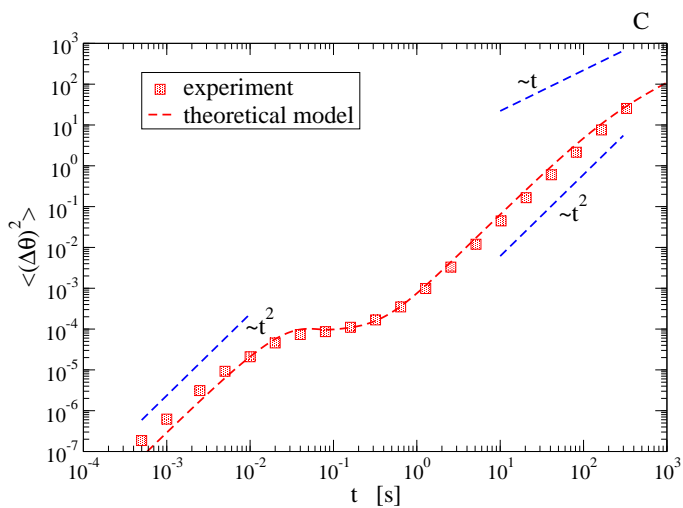}
\caption{ Comparison between the theory (dashed lines) and the 
experimental data (symbols) for the 
mean square displacement for the same case of Fig.~\ref{psd2}.}
\label{msd2}
\end{figure}

To conclude this Section, in Fig.~\ref{msd2} we show an excellent agreement for the msd
between theory and experiment in a cold liquid case. The superposition is fair along all the timescales.

\subsection{Persistency.}

\begin{figure}
\includegraphics[clip=true,angle=0,width=0.5\textwidth]{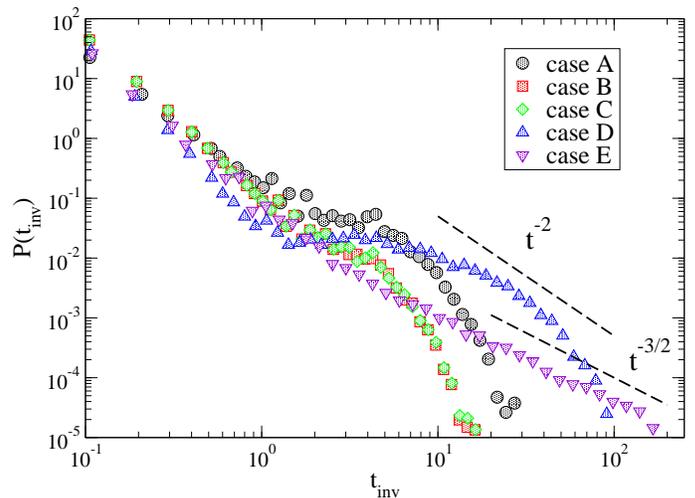}
\caption{ Distributions of inversion times for some of the cases in
  Table~\ref{casi}. All distributions present a - more or less visible
  - cut-off at a finite time. Looking at the coupled case (B-E), it is
  seen that when the cage inertia $I_0$ is increased, the cut-off time
  increases too. In the cases with largest cage inertia (D and E), the
  cut-off time is large enough to ``uncover'' some power-law decay.  
  In order to present more clearly the results, and given that cases $E$ and $F$ are very close, 
  we show only case $E$.}
\label{pers}
\end{figure} 

An intriguing counterpart of super-diffusion is a long memory in the
velocity variable $\omega(t)$. At times shorter than the typical time
needed to overcome the cage effect, the dynamics of $\omega(t)$ is
dominated by rapid intracage oscillations and fast thermal
relaxation. As seen in Fig.~\ref{autoc}, when cage inertia is large
the autocorrelation of $\omega(t)$ displays a very small long-lasting
drift. Such a weak signal in the autocorrelation
 disappears, in experiments, because of noise. A better
  observable~\cite{camille}, is the distribution of inversion times for the slow part of the signal, which is strictly
  connected to such a long memory.   In order to make contact with the
experimental results in~\cite{camille}, we have studied this
distribution in our theoretical model. To remove the effect of rapid
relaxation, we apply a filter and study
$\omega_s(t)=\frac{1}{\tau}\int_t^{t+\tau} ds \omega(s)$, using values
of $\tau=2$ larger than the cage relaxation time. An inversion time
$t_{inv}$ is the time between two zeroes of $\omega_s(t)$.  The
inversion times, therefore, represent the duration of ``persistency''
of the direction of motion.  We remark that a theoretical computation
of the distribution of persistency times is a tough task which we have
not pursued here. On the contrary, we have simulated numerically the
model~\eqref{model} and we have computed the distribution
$P(t_{inv})$, which is shown in Fig.~\ref{pers}. From this plot one
immediately appreciates the role of cage inertia $I_0$ in inducing
large inversion times. Uncoupled case ($A$) as well as the coupled
cases with zero or low inertia ($B$ and $C$) show a distribution of
$t_{inv}$ with a fast decay at a small cut-off time. When the cage
inertia $I_0$ is increased, the distribution develops a large tail
$\sim t_{inv}^{-\alpha}$ whose cut-off increases, while $\alpha \le
2$. It is interesting to notice that similar exponents have been
observed in the experimental distribution, where they were in perfect
agreement with superdiffusion predicted by a continuous time random
walk model for $\omega_s(t)$~\cite{camille}.

\subsection{Linear response}

The aim of this section is to study the response of the variable
$\omega(t)$ to the application of a force (torque) upon it. As the
model is linear, the response is always linear. However we do not
believe that the model can reproduce experimental observations when a
large force is applied, i.e. in the non-linear regime it should be
properly modified. 

The dynamical response matrix simply reads $R(t)=exp(-A t)$. Therefore
$R_{\omega\omega}(t)\overset{def}{=}R_{22}(t)=\frac{\delta
  \omega(t)}{\delta \omega(0)}$. It is interesting to notice that the
same expression is obtained by applying the Generalized
Fluctuation-Response Relation discussed in~\cite{A72,marconi2008fluctuation}:
\begin{equation}\label{gfdt}
\frac{\delta \omega(t)}{\delta \omega(0)}=-\left\langle \omega(t)\frac{\partial ln P(z,\omega,\omega_0)}{\partial \omega} |_{t=0}  \right\rangle
\end{equation}
where $P(z,\omega,\omega_0)$ is the umperturbed steady state distribution,
Eq. (\ref{disprob}) in our case. Applying the results of Sec.~\ref{statics}, we get, therefore:
\begin{equation}\label{lrex}
\frac{\delta \omega(t)}{\delta \omega(0)}= \beta_{\omega\omega}\langle \omega(t)\omega(0) \rangle +\beta_{\omega\omega_{0}}\langle \omega(t)\omega_{0}(0) \rangle+\beta_{z\omega}\langle \omega(t)z(0) \rangle,
\end{equation}
where we remind that $\beta=\sigma^{-1}$ is the inverse covariance matrix.
Such an expression makes clear the role of static correlations, which is the crucial ingredient modifying the equilibrium Fluctuation-Response relation~\cite{Villamaina2009,sar10,Crisanti2012,noiplos}

\begin{figure}
\includegraphics[clip=true,angle=0,width=0.5\textwidth]{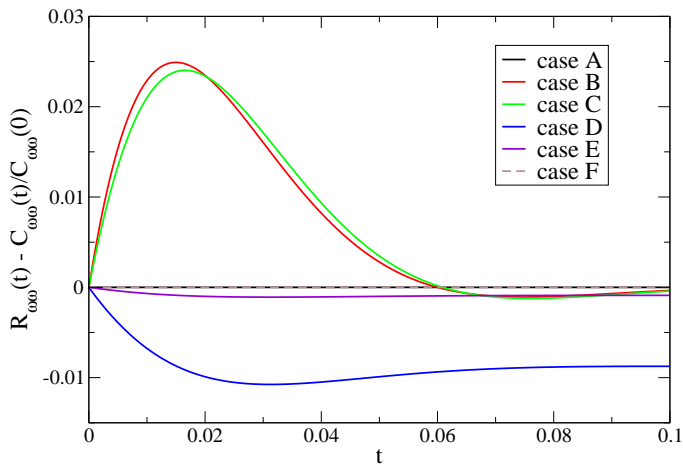}
\caption{Deviation from equilibrium measured as a distance from the
  Einstein relation, in some cases from the list in
  Table~\ref{casi}. There is no deviation in the uncoupled case (A)
  and in the coupled case when $T=T_0$ (F). All other cases show a
  deviation from the Einstein relation of similar order of magnitude,
  with the exception of the large cage inertia case (E) which is quite
  smaller.}
\label{rc}
\end{figure} 

Indeed, from expression~\eqref{lrex} and Eqs.~\eqref{covelem}, it appears that at equilibrium,
i.e. when $T=T_0$ (with or without coupling) one has
$R_{\omega\omega}(t)=C_{\omega\omega}(t)/C_{\omega\omega}(0)$, which
is a way to express the so-called Einstein relation which gives
mobility as proportional to diffusivity. 
Plots of the difference between the response $R_{\omega\omega}(t)$ and
the rescaled autocorrelation $C_{\omega\omega}(t)/C_{\omega\omega}(0)$ are shown in Fig.~\ref{rc}.
When $T \neq T_0$ (and $k>0$)
the response is always different from the rescaled autocorrelation and
the Einstein relation is always violated. An experiment comparing
linear response to autocorrelation would be able to put in evidence
the distance from equilibrium in the system.


\section{Conclusions and perspectives}

The theoretical understanding of the liquid state of granular fluids
is in an underdeveloped stage, if compared with granular gases or with
slowly tapped/sheared (or even static) granular ``solids''.  The
situation could be considered similar to the theory of liquids in the
60's, where neutron scattering spectra were explained by proposing
super-simplified models~\cite{io1,io2}.

In this spirit, here we have proposed a simple model which reproduces
two striking features of a recent experiment~\cite{camille}, namely: a
transient dynamical arrest (cage effect), and late time - almost
ballistic - superdiffusion. A qualitative comparison with the
experimental results is fair for the velocity power density spectrum,
as it includes the ``resonant'' bump associated with the cage effect
and the strong enhancement at low frequencies ($f \to 0$) which seems
related to superdiffusion. The comparison for the mean squared
displacement is even more striking, and shows how the superdiffusive
behavior emerges only when the ``cage inertia'' $I_0$ is large. In
close analogy with the experiments, also the distribution of
persistency times is strongly affected by the value of $I_0$ which
controls the cut-off time of the distribution and the possibility to
observe slow (power-law) decays. We have also discussed how the lack of thermal equilibration
between $\omega(t)$ and the collective variable $\omega_0(t)$, which
is a common feature of granular fluids due to non-conservative
interactions, could be demonstrated in future experiments through an analysis of linear response.

As a concluding remark, we wish to underline that the model proposed
is purely phenomenological and lacks a general derivation from first
principles. In particular, the experimental parameters such as the
packing fraction or the average energy input do not enter in this
simplified picture and there is no way to to deduce or estimate a
reasonable value for $I_0$.  The experimental observations suggest
that, when the ``granular temperature'' is decreased, larger and
larger values of $I_0$ are realised.  Tentative fits
  of experimental vpds at the smallest granular temperatures, with the
  formula derived here, give the values of the ``collective'' momentum
  of inertia of $I_0 \sim 10^4 I$: if we assume that this is the
  momentum of inertia of a solid made of a material density of steel
  (as the granular spheres) reduced by the packing fraction (order
  $\sim 50 \%$) it gives a radius of the order of $10^2$mm which
  matches the order of magnitude of the size of the container filled
  by grains in the experiment.  A mechanism explaining the building
up of cage inertia is still lacking and certainly deserves future
investigation.

\begin{acknowledgments}
The authors acknowledge useful discussions with A. Gnoli, A. Sarracino
and A. Vulpiani, and especially thank  A. Gnoli and C. Scalliet
for the experimetal data.
\end{acknowledgments}

%
%




\bibliography{mergedbiblio}

\end{document}